%% file: 0_main.tex
\documentclass[sigconf]{acmart}
\AtBeginDocument{%
  }

\setcopyright{acmcopyright}
\copyrightyear{2018}
\acmYear{2018}
\acmDOI{XXXXXXX.XXXXXXX}

\usepackage{color}

\usepackage[utf8]{inputenc} 
\usepackage[T1]{fontenc}    
\usepackage{hyperref}       
\usepackage{url}            
\usepackage{booktabs}       
\usepackage{amsfonts}       
\usepackage{nicefrac}       
\usepackage{microtype}      
\usepackage{xcolor}         
\usepackage{float}
\usepackage{booktabs}
\usepackage{array}
\usepackage{balance} 
\usepackage{multirow}
\usepackage{color}
\definecolor{lightgray}{RGB}{215,215,215}
\usepackage{colortbl}  
\usepackage{xcolor}
\usepackage{subfigure}
\usepackage{amsmath}

\usepackage{algorithm}  
\usepackage{amsmath}  
\usepackage{enumitem}

\floatname{algorithm}{Algorithm}

\usepackage{amsmath}  
\usepackage{enumitem}
\usepackage{amsthm}
\usepackage{bm}
\usepackage{graphicx}
\usepackage{caption}
\usepackage{wrapfig}
\usepackage{verbatim}

\newcommand{\ie}{\emph{i.e., }}
\newcommand{\eg}{\emph{e.g., }}

\copyrightyear{2025}
\acmYear{2025}
\setcopyright{acmlicensed}\acmConference[RecSys '25]{Proceedings of the Nineteenth ACM Conference on Recommender Systems}{September 22--26, 2025}{Prague, Czech Republic}
\acmBooktitle{Proceedings of the Nineteenth ACM Conference on Recommender Systems (RecSys '25), September 22--26, 2025, Prague, Czech Republic}
\acmDOI{10.1145/3705328.3748085}
\acmISBN{979-8-4007-1364-4/2025/09}
\begin{document}

\title{Heterogeneous User Modeling for LLM-based Recommendation}

\author{Honghui Bao}
\email{honghuibao2000@gmail.com}
\affiliation{
  \institution{National University of Singapore}
\country{Singapore}
}
\author{Wenjie Wang}
\email{wenjiewang96@gmail.com}
\affiliation{
\institution{University of Science and Technology of China}
\city{Hefei}
\country{China}
}

\author{Xinyu Lin}
\email{xylin1028@gmail.com}
\authornote{Corresponding author.}
\affiliation{
\institution{National University of Singapore}
\country{Singapore}
}

\author{Fengbin Zhu}
\email{zhfengbin@gmail.com}
\authornotemark[1]
\affiliation{
\institution{National University of Singapore}
\country{Singapore}
}
\author{Teng Sun}
\email{stbestforever@gmail.com}
\affiliation{
\institution{Shandong University}
\city{Qingdao}
\country{China}
}
\author{Fuli Feng}
\email{fulifeng93@gmail.com}
\orcid{0000-0002-5828-9842}
\affiliation{
\institution{University of Science and Technology of China}
\city{Hefei}
\country{China}
}

\author{Tat-Seng Chua}
\email{dcscts@nus.edu.sg}
\orcid{0000-0001-6097-7807}
\affiliation{
  \institution{National University of Singapore}
\country{Singapore}
}

\renewcommand{\shortauthors}{Honghui Bao et al.}


\begin{abstract}

Leveraging Large Language Models (LLMs) for recommendation has demonstrated notable success in various domains, showcasing their potential for open-domain recommendation.
A key challenge to advancing open-domain recommendation lies in effectively modeling user preferences from users' heterogeneous behaviors across multiple domains.
Existing approaches, including ID-based and semantic-based modeling, struggle with poor generalization, an inability to compress noisy interactions effectively, and the domain seesaw phenomenon.
To address these challenges, 
we propose a Heterogeneous User Modeling (HUM) method, which incorporates a compression enhancer and a robustness enhancer for LLM-based recommendation. 
The compression enhancer uses a customized prompt to compress heterogeneous behaviors into a tailored token, while a masking mechanism enhances cross-domain knowledge extraction and understanding.
The robustness enhancer introduces a domain importance score to mitigate the domain seesaw phenomenon by guiding domain optimization. 
Extensive experiments on heterogeneous datasets validate that HUM effectively models user heterogeneity by achieving both high efficacy and robustness, leading to superior performance in open-domain recommendation.

\end{abstract}





\maketitle

\input{1_intro}
\input{2_0_Task}

\input{2_method}
\input{3_exp}
\input{4_related_work}
\input{5_conclusion}

\clearpage

{
\balance
\bibliographystyle{ACM-Reference-Format}
\bibliography{reference}
}



\end{document}

%% file: 1_intro.tex
\section{Introduction}


\begin{figure}[t]
\setlength{\abovecaptionskip}{0cm}
\setlength{\belowcaptionskip}{-0.30cm}
\centering
\includegraphics[scale=0.73]{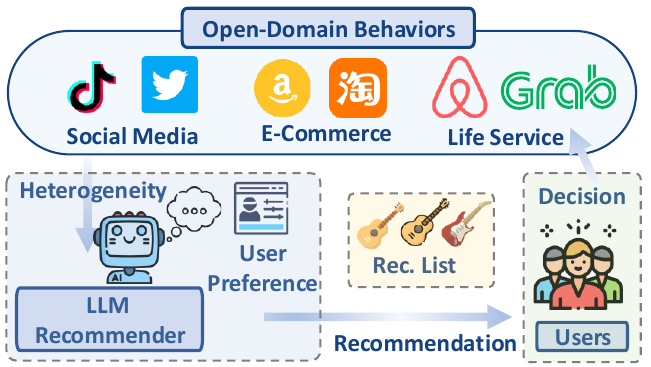}
\caption{Overview of heterogeneous user modeling in open-domain environments.}
\label{fig:overview}
\end{figure}
Utilizing Large Language Models (LLMs) for recommendation has achieved remarkable results across various domains, including Games~\cite{zheng2023adapting}, Movie~\cite{lin2024data}, and Toys~\cite{tan2024towards}.
LLMs' rich world knowledge and powerful capabilities (\eg reasoning and generalization) reveal potential to achieve an ambitious objective: uncovering implicit preferences in open-domain user behaviors to deliver personalized recommendations, as illustrated in Figure~\ref{fig:overview}.
The key to achieving this objective lies in leveraging LLMs to model user preferences from their heterogeneous behaviors across multiple domains.

A closely related line of research to modeling user preferences from heterogeneous behaviors is multi-domain recommendation~\cite{cao2023unicdr, park2024pacer, hou2022unisrec}. 
Typically, given a user's multi-domain interactions, multi-domain recommendation learns a user representation to model user preference for recommendation across various domains.
Existing work can be categorized into two types: 
\begin{itemize}[leftmargin=*]
    \item {ID-based modeling}~\cite{cao2023unicdr,park2024pacer} uses unique identifiers (IDs) to obtain embeddings and incorporates multi-domain interactions to learn personalized user representations. 
    However, this approach heavily relies on substantial interactions, which can result in poor generalization in cold-start scenarios~\cite{fu2023unified, tang2023one}.
    \item Semantic-based modeling~\cite{hou2022unisrec,Li2023TextIA, tang2023one} learns user representations from the semantic information of interacted items across domains, which can alleviate the cold-start generalization issue.
    Existing methods usually utilize various language models (\eg LongFormer~\cite{Li2023TextIA}, BERT~\cite{tang2023one}, T5~\cite{tan2024towards}) to encode the user history to obtain user representations. 
    The issue with this approach is twofold: 
First, heterogeneous user interactions often contain noise (\ie interactions irrelevant to or conflicting with the target domain's recommendation). Effectively compressing such noisy behaviors is crucial for learning accurate user representations in multi-domain recommendation. However, existing semantic-based methods~\cite{Li2023TextIA,tang2023one} typically extract the overall semantics without being explicitly trained to handle user heterogeneity. As a result, they lack the ability to filter out noisy signals, which leads to representations misaligned with users' true preferences and ultimately degrades recommendation performance.
Second, previous models tend to favor specific domains during training, leading to the domain seesaw phenomenon~\cite{fu2023unified}, where improvements in one domain may result in a decline in another.

\end{itemize}

To overcome the above issues, we summarize two principal objectives for heterogeneous user modeling based on LLMs:
1) Strong compression capability, which enables LLMs to understand user's heterogeneous behaviors by excluding noise and to extract representations by compressing user preferences.
2) Strong robustness, which ensures that domains are optimized without favoring a specific domain, alleviating the domain seesaw phenomenon and facilitating accurate recommendations.

\begin{figure}[t]
\setlength{\belowcaptionskip}{-0.50cm}
\centering
\includegraphics[scale=0.375]{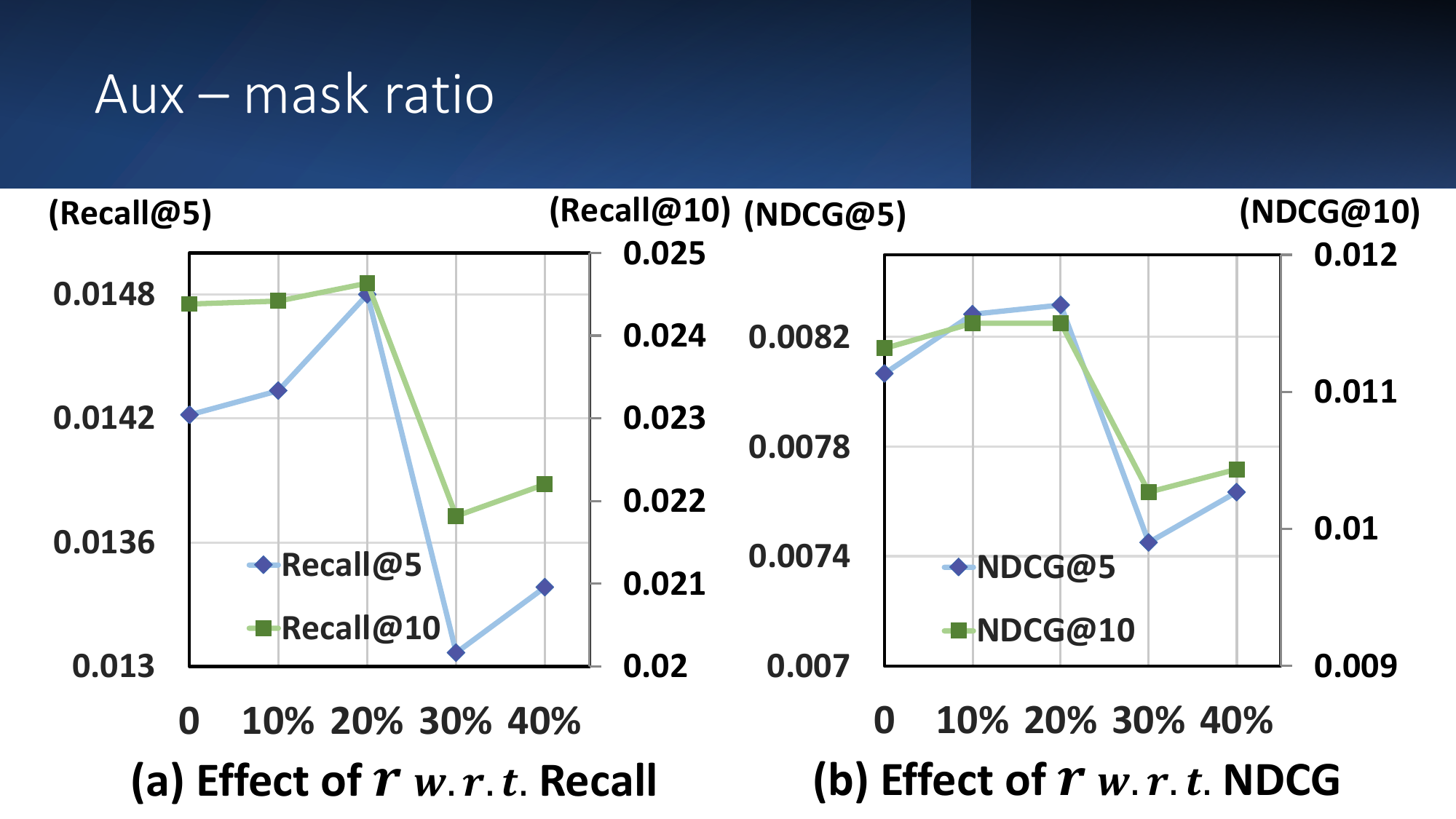}
\caption{Illustration of the incapability of directly utilizing LLMs to learn heterogeneity and the domain seesaw phenomenon in three domains. ``h.'' denotes ``heterogeneity''.}
\label{fig:domain_seesaw}
\end{figure}

A straightforward approach is to directly utilize LLMs for heterogeneous user modeling, as LLMs possess extensive world knowledge and strong reasoning abilities. Ideally, LLMs could understand heterogeneous textual user interactions to generate excellent representations. However, this intuitive approach fails to achieve the two objectives above due to two reasons:
1) It struggles to effectively distinguish heterogeneous information, as evidenced by LLM-Rec-Qwen, which directly utilizes LLMs, as shown in Figure~\ref{fig:domain_seesaw}(a), where incorporating heterogeneous interactions inversely hurt the model’s performance.
2) It still suffers from the domain seesaw phenomenon, as illustrated in Figure~\ref{fig:domain_seesaw}(b), where the improvement in the Auto domain's performance leads to a decline in the performance of the Scientific domain and the Office domain.

To this end, we propose a \textbf{H}eterogeneous \textbf{U}ser \textbf{M}odeling approach for LLM-based Recommendation named HUM. 
HUM incorporates two kinds of enhancer to promote the heterogeneous user modeling. 
In particular, the compression enhancer first employs a unidirectional decoder model (\eg Qwen) to compress users' heterogeneous interactions into a tailored user token, guided by a compression prompt. Moreover, this enhancer leverages a masking mechanism to enhance the extraction of transferable knowledge across domains in multi-domain interactions, thereby improving LLMs' ability to understand heterogeneous user behaviors.
Besides, robustness enhancer introduces a domain importance score to balance domains' optimization which aim to mitigate domain seesaw phenomenon.
Extensive experiments on multi-domain datasets with in-depth investigation towards robustness on seen domains, noise resistance, generalization on unseen domains, scalability have validated that HUM can achieve superior heterogeneous user modeling by simultaneously considering strong compression capability and robustness. 
For reproducibility, we release our code and data at \url{https://github.com/HonghuiBao2000/HUM}.

To sum up, the contributions of this work are as follows.
\begin{itemize}[leftmargin=*]
    \item We highlight the significance of heterogeneous user modeling in open-domain recommendation and comprehensively analyze the essential features of heterogeneous user modeling. 
    \item We propose a novel heterogeneous user modeling method, HUM, which leverages a compression enhancer and a robustness enhancer to achieve strong compression capabilities for LLMs and high robustness across diverse domains.

    \item We conduct extensive experiments on heterogeneous datasets, coupled with the in-depth investigation with diverse settings, validating that HUM outperforms existing heterogeneous user modeling methods for LLM-based recommendation.
    
\end{itemize}

%% file: 2_0_Task.tex
\section{Preliminaries}
\label{sec:task}

\vspace{2pt}
\noindent$\bullet\quad$\textbf{Multi-domain sequential recommendation}.
Multi-domain recommendation \cite{tang2023one, park2024pacer} captures user preferences from heterogeneous interactions and recommends items across multiple domains.
Formally, we define a total of \( N \) domains, namely 
$\mathcal{D}=\left\{d_1, d_2, \ldots, d_{N}\right\}
$ in a multi-domain environment. For each domain $d_i$, we have the sets \( \{ \mathcal{U}_i, \mathcal{V}_i, \mathcal{I}_i \} \), which denotes user, item, and interaction. Then, we can define user set, item set, and interactions set of this environment as:
$\mathcal{U} = \bigcup_{i=1}^{N} \mathcal{U}_i$, $\mathcal{V} = \bigcup_{i=1}^{N} \mathcal{V}_i$, $ \mathcal{I} = \bigcup_{i=1}^{N} \mathcal{I}_i$.
We denote the user's heterogeneous interaction sequence in chronological order as 
\begin{equation}
    u_i = (v_1, v_2, v_3, \dots, v_{L}),
\end{equation}
where each interacted item $v_i$ belongs to a specific domain, and $L$ denotes the length of the user’s sequence. 
Then, given a user's heterogeneous interaction sequence and a target domain, multi-domain sequential recommendation aims to model the heterogeneous user preference to predict the next item for the target domain.

\begin{figure}[]
\setlength{\abovecaptionskip}{0.01cm}
\setlength{\belowcaptionskip}{-0.50cm}
\centering
\includegraphics[scale=0.63]{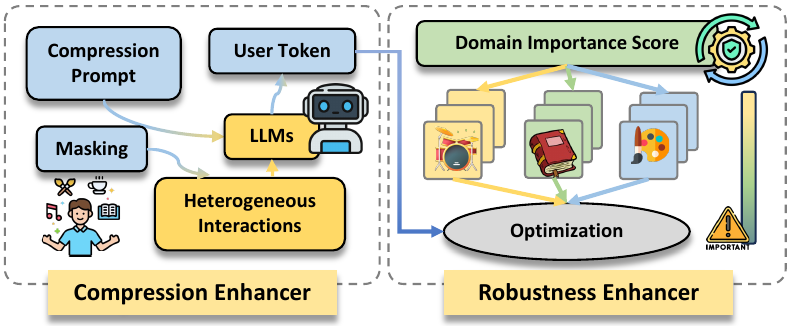}
\caption{Overview of HUM. HUM first employs a compression enhancer to compress heterogeneous user behaviors. Then, it applies a robustness enhancer to mitigate domain conflicts (\ie the domain seesaw phenomenon).}
\label{fig:method}
\end{figure}

\vspace{2pt}
\noindent$\bullet\quad$\textbf{Heterogeneous user modeling}.
\label{sec:task_def}
The key to multi-domain sequential recommendation is effectively capturing latent user preferences from noisy heterogeneous interactions, enabling accurate recommendations in the target domain.
In previous heterogeneous user modeling approaches, ID-based methods index user interactions using unique IDs to obtain representations.
Semantic-based methods leverage semantic tokens, derived from item descriptions or titles, to obtain representations.
In this paper, we focus on the semantic-based approach because it offers several advantages for multi-domain recommendation:
1) it enables better generalization to unseen scenarios(\eg cold users and items).
2) semantic representations facilitate cross-domain adaptability, enabling the transfer of user preferences across contexts, even in the absence of overlapping users, making them particularly effective in multi-domain settings.

Based on the semantic-based approach, each item is represented by its textual title, \ie $t$, and the user interaction sequence is transformed into a title sequence, \ie $u_i = (t_{v_1}, t_{v_2}, t_{v_3}, \dots, t_{v_{L}})$.  
Therefore, semantic-based user modeling methods use the textual representations of users and items as input, feeding them into LLMs to extract specific token hidden vectors (\eg\texttt{[EOS]}~\cite{tang2023one}) from the last layer as the user and item representations. By calculating the inner product between the extracted user and item representations, we measure the similarity between them. The items with the highest scores from the sorted list are then selected as recommended items. 

However, existing semantic-based methods~\cite{tang2023one,Li2023TextIA} face two issues:
First, they rely on language models to understand and summarize the overall semantics of the heterogeneous interactions. While effective for general text modeling, this approach lacks the capacity to focus on fine-grained differences within heterogeneous behaviors. As a result, when faced with noisy interactions, the model tends to capture an averaged or diluted semantic representation, rather than the user's true preference towards the target domain.
Second, during multi-domain training, models often favor domains with richer or more informative data, as they contribute more strongly to the overall loss reduction. This domain imbalance leads to the domain seesaw phenomenon~\cite{fu2023unified}, where improving performance in one domain comes at the cost of degrading another. 
Motivated by these observations, we argue that heterogeneous user modeling requires: (1) strong compression capability that enables the model to selectively identify and compress the information most relevant to the user's true preferences; and (2) strong robustness that prevents overfitting to dominant domains, allowing for balanced optimization across all domains.


%% file: 2_method.tex
\section{Method}
\label{sec:method}

In order to pursue effective heterogeneous user modeling, we propose HUM, which includes two key components, \ie the compression enhancer, designed to improve LLMs' compression capabilities for noisy heterogeneous interactions, and a robustness enhancer to pursue a stable performance across domains. The overview of our method is presented in Figure~\ref{fig:method}.

\subsection{Compression Enhancer}

\begin{figure}[]
\setlength{\abovecaptionskip}{0.01cm}
\setlength{\belowcaptionskip}{-0.50cm}
\centering
\includegraphics[scale=0.47]{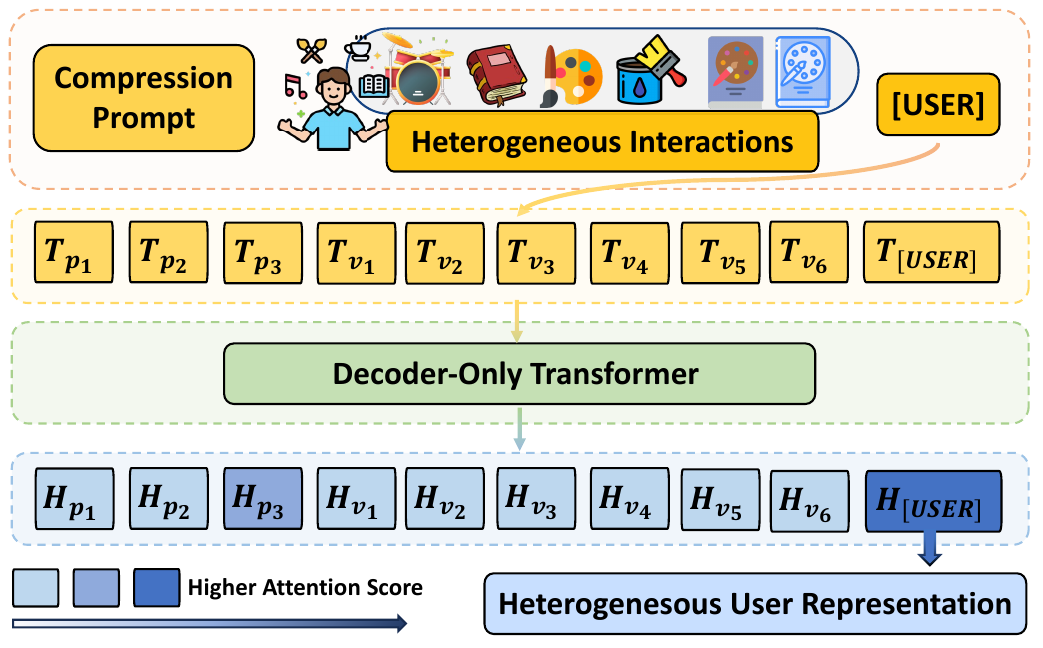}
\caption{Illustration of compression, which moves from input to model to representation, where ``T'' represents a token, ``H'' represents token's last-layer hidden vector.}
\label{fig:method_compression}
\end{figure}

The key to achieving multi-domain recommendation lies in leveraging LLMs to model user preferences within their heterogeneous multi-domain interactions. 
However, directly utilizing LLMs fails to accurately extract user representations from heterogeneous information, as this approach neither accurately infers user preferences from noisy heterogeneous data nor effectively compresses these interactions into meaningful user representations.
To address this challenge, compression enhancer guides LLMs to compress user interactions via a compression prompt and uses a tailored user token to represent heterogeneous user preferences, as shown in Figure~\ref{fig:method_compression}. In addition, it employs a masking mechanism to facilitate information fusion while minimizing the impact of noisy interactions.


\vspace{2pt}
\noindent$\bullet\quad$\textbf{Compression prompt}.
LLMs possess extensive world knowledge and demonstrate remarkable capabilities in generative tasks~\cite{brown2020languagemodelsfewshotlearners}. However, as shown in Figure~\ref{fig:domain_seesaw}, directly applying LLMs to heterogeneous user modeling yields suboptimal performance, largely due to their lack of structural guidance in compressing diverse user behaviors.
Given LLMs' strong instruction-following abilities~\cite{sanh2021multitask}, we propose to explicitly leverage this characteristic for representation compression. To this end, we design a simple yet effective compression prompt:  \textbf{``Compress the following description about the user or item into the last token:''}.
This prompt guides the LLM to utilize its world knowledge and reasoning capabilities to condense useful information from heterogeneous user interactions into a compact representation~\cite{deletang2023language}. 
Empirically, this prompt-enhanced fine-tuning approach improves the LLM’s ability to handle heterogeneous inputs and generates more accurate user representations across domains (see Section~\ref{ablation_sec}).

\vspace{2pt}
\noindent$\bullet\quad$\textbf{User token}.
To obtain user representations from LLMs, existing methods~\cite{neelakantan2022text, tang2023one} typically adopt the hidden state of the final \texttt{[EOS]} token as the user representation. However, unlike special-purpose tokens such as \texttt{[CLS]}~\cite{devlin2018bert}, the \texttt{[EOS]} token is pretrained as a generic termination signal, not as an information aggregator. Consequently, using it as a representation carrier during fine-tuning may introduce noise or interfere with information integration, especially in tasks involving heterogeneous user interactions.
To address this, we introduce a special token \textbf{\texttt{[USER]}}, explicitly inserted after the user input to serve as a designated aggregation anchor. This token is trained to compress and carry semantically meaningful information from heterogeneous behavior sequences.
Empirically, this design substantially improves representation quality, demonstrating both strong in-domain performance and enhanced generalization across unseen domains (see Section~\ref{sec: generalization}).

\vspace{2pt}
\noindent$\bullet\quad$\textbf{Model input}.
In HUM, each item $v$ is represented by its title $t$, and a user by the concatenation of interacted items' title sequences. We then adopt a compression prompt and tailored user token in the sequence. The model inputs for users are denoted as:
\begin{equation}\label{eqn:input}
\begin{aligned}
X_u = \{prompt, t_1, t_2, t_3, \dots, t_L,\textbf{\texttt{[USER]}}\}
\end{aligned}
\end{equation}
where $X_u$ is a token sequence containing the compression prompt, titles of historically interacted items, and a tailored user token.
Each item is treated as a special case of a user who interacted with only one item, and its model input is constructed as:
\begin{equation}\label{eqn:item_input}
\begin{aligned}
X_{v} = \{prompt, t_v,\textbf{\texttt{[USER]}}\}
\end{aligned}
\end{equation}


\vspace{2pt}
\noindent$\bullet\quad$\textbf{Compression optimization}.
During training, we feed both user and item inputs into a single LLM for joint learning. Specifically, we extract the hidden vector of the tailored user token from the last hidden layer as the user representation and item representation, which can be formulated as:
\begin{equation}
\begin{aligned}
\bm{e}_u = \text{LLM}(X_u),\quad   \bm{e}_{v} = \text{LLM}(X_{v})
\end{aligned}
\end{equation}

where $\bm{e}_u$ and $\bm{e}_{v}$ are the representation of user and item. 
To measure the relevance between users and items, we compute similarity between user and item representations using inner product. Furthermore, to improve the model's discriminative ability between the target item and its similar items, we sample $N$ negative items from the same domain for each target item and adopt contrastive learning to optimize the model. Formally, we have:
\begin{equation}
    \mathcal{L}=-\sum_{i=1}^{B}\frac{\exp (<\bm{e}_{u}^{i}, \bm{e}_{v_{+}}>)}{\exp (<\bm{e}_{u}^{i}, \bm{e}_{v_{+}}>) + \sum_{j=1}^{N}\exp(<\bm{e}_{u}^{i},\bm{e}_{v_{j}}>)}
\end{equation}
where $\bm{e}_{u}^{i}$ is the user representation, $\bm{e}_{v_{+}}$ denotes the ground truth item representation and $\bm{e}_{j\in{\{1, \dots, N\}}}$ represents the representation of all negative samples, $<\cdot ,\cdot>$ denotes the inner product, and $B$ is the batch size.
This contrastive objective not only encourages the model to align the user representation with the correct item, but also implicitly guides the model to distinguish useful information from noisy heterogeneous user interactions. As a result, the model gradually learns to focus on domain-relevant patterns that are critical for accurate recommendation in the target domain.

\vspace{2pt}
\noindent$\bullet\quad$\textbf{Masking mechanism}.
To improve the quality of user representations for the target domain, it is crucial to distinguish informative signals from noisy behaviors in heterogeneous interactions.
To this end, we propose a masking mechanism that randomly masks a portion of target-domain items in the user history with a probability $r$, \ie, the mask ratio. The masked input can then be represented as: \begin{equation}\label{eqn:revised_input} \begin{aligned} X'_u \leftarrow \text{Mask}(X_u, r) \end{aligned} \end{equation} where $X_u$ is the original user history and $X'_u$ is the masked version.
During training, the model is optimized on the masked inputs. By partially removing the target-domain items, the model is encouraged to identify and leverage helpful context from other domains, thus learning to discriminate informative behaviors from noisy ones without explicit supervision.

\subsection{Robustness Enhancer}
\begin{figure}[] \setlength{\abovecaptionskip}{0.01cm} \setlength{\belowcaptionskip}{-0.50cm} \centering \includegraphics[scale=0.7]{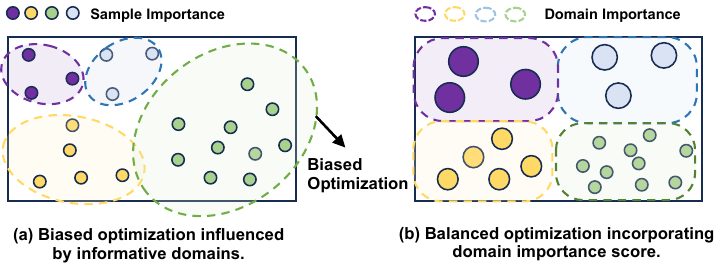} \caption{Illustration of biased optimization dominated by informative domains and balanced optimization incorporating domain importance.} \label{fig:method_robust} \end{figure}

Incorporating more interaction data facilitates the construction of more comprehensive user representations. However, scaling up the dataset also introduces increased heterogeneity across sources, \ie different domains~\cite{iacob2024deptdecoupledembeddingspretraining}.
This heterogeneity biases the optimization process, allowing domains with more informative samples to dominate training, which leads to the domain seesaw phenomenon~\cite{fu2023unified}, as illustrated in Figure~\ref{fig:method_robust}(a).
To address this, we propose a domain importance mechanism that quantifies each domain's importance during training, and further apply a domain smoothing strategy to ensure training stability.

\vspace{2pt} \noindent$\bullet\quad$\textbf{Domain importance.}
To mitigate the domain seesaw phenomenon, we introduce a Domain Importance (DI) score that guides the model toward a more balanced optimization across domains.
Specifically, we define domain importance based on empirical risk, \ie the average training loss per domain: \begin{equation}\label{eqn:di} \begin{aligned} DI_i = \frac{\mathcal{L}\left(d_i, \theta\right)}{\sum_j \mathcal{L}\left(d_j, \theta\right)} = \frac{\sum_{s_{ik} \in d_i} \mathcal{L}\left(s_{ik}, \theta\right)}{\sum_{s_{jl} \in \mathbb{D}} \mathcal{L}\left(s_{jl}, \theta\right)} \end{aligned} \end{equation} Here, $DI_i$ denotes the importance of domain $d_i$, $s_{ik}$ is the $k$-th sample in $d_i$, and $\mathbb{D}$ is the set of all training samples. $\theta$ represents either the full model parameters or parameters trained via efficient methods such as LoRA~\cite{hu2021lora}.
Intuitively, a higher $DI$ indicates higher empirical risk, suggesting the domain is under-optimized and should be prioritized in subsequent training steps.
This score thus serves as a domain-wise weighting factor to regularize the optimization process, encouraging the model to focus more on domains that require additional learning.

\vspace{2pt} \noindent$\bullet\quad$\textbf{Domain smoothing.}
Directly applying domain importance during training may cause instability, particularly for sparse domains with few samples~\cite{10.1145/3485447.3512255}. Due to the scarcity of training samples, the estimated losses for these domains can exhibit high variance, leading to noisy or unreliable $DI$ values that do not accurately reflect their true importance.
To alleviate this, we adopt a domain smoothing strategy following~\cite{10.1145/3485447.3512255, lin2024temporally, zhao2023popularity}, which incorporates prior information into the current domain importance estimation.
We maintain a domain importance vector $\bm{w}$ and introduce a regulating factor $\alpha$ to control the update smoothness. This leads to the following optimization objective:  \begin{equation}
\small
\label{eqn:derive_kkt}
\begin{aligned}
&\underset{w}{\min} \sum_i w_i \mathcal{L} \left(d_i, \theta\right) + \frac{1}{\alpha} \text{KL}(\bm{w}, \bm{w}^{t-1}) \\
= 
&\underset{w}{\min} \sum_i w_i \mathcal{L} \left(d_i, \theta\right) + \frac{1}{\alpha} \sum_i w_i \log \frac{w_i}{w_i^{t-1}}, \\
&\mathrm{s.t.} \begin{cases}  
\sum_i w_i = 1, \\  
w_i > 0 
\end{cases}
\end{aligned}
\end{equation}
where $t$ denotes  current update step, and $\text{KL}(\bm{w},\bm{w}^{t-1})$ is the KL divergence between the current and previous weights.
Following~\cite{lin2024temporally}, we apply KKT conditions to obtain the closed-form solution: \begin{equation} w_i^t = \frac{w_i^{t-1} \cdot \exp\left(\alpha \cdot \mathcal{L}(d_i, \theta)\right)}{\sum_j w_j^{t-1} \cdot \exp\left(\alpha \cdot \mathcal{L}(d_j, \theta)\right)} \end{equation}
The final training loss becomes: \begin{equation} \mathcal{L}_{\text{HUM}} = \sum_i w_i \cdot \mathcal{L}(d_i, \theta) \end{equation}
By dynamically adjusting domain contributions during training, this mechanism enables the model to concentrate more effectively on under-optimized yet informative domains, while reducing the dominance of over-represented ones.
As shown in Figure~\ref{fig:method_robust}(b), this strategy enhances both training stability and the model's ability to robustly capture cross-domain knowledge.

\vspace{2pt}
\noindent$\bullet\quad$\textbf{HUM Inference}. 
At inference, we deactivate the masking mechanism and generate the heterogeneous representations of users as well as the representations of all items. Then, given a user heterogeneous interaction sequence and a target domain, we compute the relevance score between the user's representation $\bm{e}_u$ and the representations $\bm{e}_v$ (\ie $<\bm{e}_u, \bm{e}_v>)$ of all items within the target domain. 
Finally, we can obtain the recommendation list for the target domain by ranking the relevance scores. 

%% file: 3_exp.tex
\section{Experiments}
\label{sec:exp}

\begin{table*}[]
\setlength{\abovecaptionskip}{0.3cm}
\setlength{\belowcaptionskip}{0.2cm}
\caption{Overall performance comparison between the baselines and HUM on six-domain datasets. For each domain, the bold results highlight the best results, while the second-best ones are underlined.}
\label{tab:performance_comparison}

\setlength{\tabcolsep}{1.95mm}{
\resizebox{0.88\textwidth}{!}{%

\begin{tabular}{ccccccccccc}
\toprule
\multicolumn{1}{l}{}                                      & \multicolumn{1}{l}{} & \multicolumn{4}{c}{\textbf{ID-based Method}}         & \multicolumn{5}{c}{\textbf{Semantic-based Method}}                                                                 \\ 
\cmidrule(lr){3-6} \cmidrule(lr){7-11} 
 
\multicolumn{1}{l}{\textbf{Dataset}}                      & \textbf{Metric}      & \textbf{SASRec} & \textbf{STAR} & \textbf{SyNCRec} & \textbf{UniCDR} & \textbf{IDGenRec} & \textbf{UniSRec} & \textbf{Recformer} & \textbf{LLM-Rec} & \textbf{HUM}                        \\ \midrule \midrule
\multicolumn{1}{c|}{\multirow{4}{*}{\textbf{Books}}}      & \textbf{R@5}         & 0.0034          & 0.0005        & 0.0005           & 0.0050          & 0.0039            & 0.0177           & {\underline {0.0374}}       & 0.0328           & \textbf{0.0432}                     \\
\multicolumn{1}{c|}{}                                     & \textbf{R@10}        & 0.0036          & 0.0007        & 0.0007           & 0.0050          & 0.0065            & 0.0308           & {\underline {0.0513}}      & 0.0497           & \textbf{0.0667}                     \\
\multicolumn{1}{c|}{}                                     & \textbf{N@5}         & 0.0024          & 0.0003        & 0.0003           & 0.0025          & 0.0025            & 0.0105           & {\underline {0.0216}}       & 0.0189           & \textbf{0.0247}                     \\
\multicolumn{1}{c|}{}                                     & \textbf{N@10}        & 0.0025          & 0.0004        & 0.0004           & 0.0025          & 0.0034            & 0.0146           & {\underline {0.0261}}       & 0.0243           & \textbf{0.0322}                     \\ \midrule
\multicolumn{1}{c|}{\multirow{4}{*}{\textbf{Office}}}     & \textbf{R@5}         & 0.0012          & 0.0028        & 0.0011           & 0.0020          & 0.0027            & 0.0048           & {\underline {0.0079}}       & 0.0071           & \textbf{0.0091}                     \\
\multicolumn{1}{c|}{}                                     & \textbf{R@10}        & 0.0017          & 0.0034        & 0.0022           & 0.0026          & 0.0034            & 0.0081           & {\underline {0.0136}}       & 0.0120           & \textbf{0.0155}                     \\
\multicolumn{1}{c|}{}                                     & \textbf{N@5}         & 0.0007          & 0.0019        & 0.0008           & 0.0005          & 0.0020            & 0.0030           & {\underline {0.0044}}       & 0.0043           & \textbf{0.0054}                     \\
\multicolumn{1}{c|}{}                                     & \textbf{N@10}        & 0.0009          & 0.0021        & 0.0012           & 0.0005          & 0.0022            & 0.0040           & {\underline {0.0062}}       & 0.0059           & \textbf{0.0074}                     \\ \midrule
\multicolumn{1}{c|}{\multirow{4}{*}{\textbf{Scientific}}} & \textbf{R@5}         & 0.0053          & 0.0054        & 0.0036           & 0.0110          & 0.0049            & 0.0067           & 0.0106             & {\underline { 0.0143}}     & \textbf{0.0145}                     \\
\multicolumn{1}{c|}{}                                     & \textbf{R@10}        & 0.0085          & 0.0074        & 0.0048           & 0.0160          & 0.0076            & 0.0109           & 0.0191             & {\underline { 0.0238}}     & \textbf{0.0246}                     \\
\multicolumn{1}{c|}{}                                     & \textbf{N@5}         & 0.0032          & 0.0033        & 0.0025           & 0.0042          & 0.0032            & 0.0040           & 0.0055             & {\underline { 0.0079}}     & \textbf{0.0086}                     \\
\multicolumn{1}{c|}{}                                     & \textbf{N@10}        & 0.0042          & 0.0040        & 0.0029           & 0.0050          & 0.0041            & 0.0053           & 0.0082             & {\underline { 0.0110}}     & \multicolumn{1}{l}{\textbf{0.0119}} \\ \midrule
\multicolumn{1}{c|}{\multirow{4}{*}{\textbf{CDs}}}        & \textbf{R@5}         & 0.0055          & 0.0000        & 0.0000           & 0.0031          & 0.0043            & 0.0056           & {\underline { 0.0075}}       & 0.0074           & \textbf{0.0099}                     \\
\multicolumn{1}{c|}{}                                     & \textbf{R@10}        & 0.0091          & 0.0000        & 0.0000           & 0.0031          & 0.0050            & 0.0093           & 0.0121             & {\underline { 0.0143}}     & \textbf{0.0161}                     \\
\multicolumn{1}{c|}{}                                     & \textbf{N@5}         & 0.0037          & 0.0000        & 0.0000           & 0.0008          & 0.0036            & 0.0034           & {\underline { 0.0045}}       & 0.0042           & \textbf{0.0049}                     \\
\multicolumn{1}{c|}{}                                     & \textbf{N@10}        & 0.0050          & 0.0000        & 0.0000           & 0.0008          & 0.0039            & 0.0046           & 0.0060             & {\underline { 0.0064}}     & \textbf{0.0069}                     \\ \midrule
\multicolumn{1}{c|}{\multirow{4}{*}{\textbf{Auto}}}       & \textbf{R@5}         & 0.0013          & 0.0014        & 0.0008           & 0.0013          & 0.0012            & 0.0016           & {\underline { 0.0075}}       & 0.0057           & \textbf{0.0080}                     \\
\multicolumn{1}{c|}{}                                     & \textbf{R@10}        & 0.0018          & 0.0029        & 0.0025           & 0.0019          & 0.0018            & 0.0056           & {\underline { 0.0116}}       & 0.0106           & \textbf{0.0153}                     \\
\multicolumn{1}{c|}{}                                     & \textbf{N@5}         & 0.0010          & 0.0008        & 0.0005           & 0.0007          & 0.0007            & 0.0007           & \textbf{0.0044}    & 0.0032           & \textbf{0.0044}                     \\
\multicolumn{1}{c|}{}                                     & \textbf{N@10}        & 0.0012          & 0.0012        & 0.0010           & 0.0008          & 0.0009            & 0.0021           & {\underline { 0.0057}}       & 0.0048           & \textbf{0.0067}                     \\ \midrule
\multicolumn{1}{c|}{\multirow{4}{*}{\textbf{Tools}}}      & \textbf{R@5}         & 0.0003          & 0.0006        & 0.0003           & 0.0020          & 0.0004            & 0.0021           & 0.0038             & {\underline { 0.0050}}     & \textbf{0.0052}                     \\
\multicolumn{1}{c|}{}                                     & \textbf{R@10}        & 0.0009          & 0.0011        & 0.0008           & 0.0033          & 0.0009            & 0.0042           & {\underline { 0.0078}}       & 0.0074           & \textbf{0.0098}                     \\
\multicolumn{1}{c|}{}                                     & \textbf{N@5}         & 0.0002          & 0.0003        & 0.0003           & 0.0009          & 0.0002            & 0.0012           & 0.0021             & \textbf{0.0031}  & {\underline { 0.0029}}                        \\
\multicolumn{1}{c|}{}                                     & \textbf{N@10}        & 0.0003          & 0.0005        & 0.0004           & 0.0011          & 0.0004            & 0.0019           & 0.0034             & {\underline { 0.0039}}     & \textbf{0.0044}                     \\ 

\bottomrule  
\end{tabular}
}}
\end{table*}

\begin{itemize}[leftmargin=*]
    \item \textbf{RQ1:} How does our proposed HUM perform compared to different kinds of heterogeneous user modeling methods?
    \item \textbf{RQ2:} How do the components of HUM (\eg compression prompt, user token, masking mechanism and domain importance score) affect the performance? 
    \item \textbf{RQ3:} How does HUM demonstrate its potential in open-domain recommendation through generalization and scalability?
\end{itemize}


\subsection{Experimental Settings}

\subsubsection{\textbf{Datasets}} 
To evaluate the performance of HUM in heterogeneous setting, we conduct empirical experiments on six different domains of up-to-date Amazon review datasets~\cite{hou2024bridging}, namely
``Scientific", ``Office", ``Books", ``CDs", ``Auto", ``Tools".
In particular, to ensure heterogeneity, we calculate the pairwise similarity between all domains in Amazon review datasets~\cite{hou2024bridging} and select subsets (\ie the 6 domains) that include some domains with strong similarities and others that exhibit notable differences.
We adopt the pre-processing techniques from previous work~\cite{tang2023one}, discarding sparse users and items with interactions less than 5.
We sample 10000 users for each domain, and collect historical interaction behaviors for each user and arrange them in chronological order to construct sequences.
To simulate real-world recommendation scenarios, we use two absolute timestamps from Amazon review datasets~\cite{hou2024bridging} to split the dataset.
For training, we follow~\cite{tang2023one} to restrict the number of items in a user's history to 10.

\subsubsection{\textbf{Baselines}}

We compare our method with competitive baselines within two groups of work, ID-based modeling methods and semantic-based modeling methods:
1) \textbf{SASRec}~\cite{kang2018self} employs self-attention mechanisms to model long-term dependencies in user interaction history.
2) \textbf{STAR}~\cite{sheng2021one} employs a shared central network for all domains and domain-specific networks tailored to each domain to model user interactions.
3) \textbf{SyNCRec}~\cite{park2024pacer} proposes an asymmetric cooperative network to decouple domain-hybrid user modeling and domain-specific user modeling, aiming to alleviate negative transfer.
4) \textbf{UniCDR}~\cite{cao2023unicdr} provide a unified framework to universally model different CDR scenarios by transferring the domain-shared information.
5) \textbf{IDGenRec}~\cite{tan2024towards} utilize semantic identifiers created from textual ID generator to represent user in generative recommendation.
6) \textbf{UniSRec}~\cite{hou2022unisrec} uses description of items and users to learn transferable representations across different domains.
7) \textbf{Recformer}~\cite{Li2023TextIA} represent an item as a word sequence by flattening its key-value attributes described in text, making a user's item sequence as a sequence of sequences.
8) \textbf{LLM-Rec}~\cite{tang2023one} adopt descriptive information of users' mixed sequence from multi-domain to build universal representation via LLMs. Here, we default to using BERT~\cite{devlin2018bert} as the backbone, since it demonstrated the best performance in the original paper.
\subsubsection{\textbf{Evaluation Settings}} 

To evaluate the performance, we adopt two widely used metrics Recall@K and NDCG@K, where K is set to 5 and 10. For all domains, we adopt a full-ranking strategy~\cite{Li2023TextIA} for evaluation, \ie ranking the ground-truth item of each user sequence among all items in the target domain.
\subsubsection{\textbf{Implementation Details}} 

We build HUM based on a representative LLM, \ie Qwen2.5-1.5b~\cite{qwen2.5}. 
Moreover, we perform the parameter-efficient fine-tuning technique 
LoRA~\cite{hu2021lora} to fine-tune Qwen2.5-1.5b.
All the experiments are conducted on 4 NVIDIA RTX A5000 GPUs. Besides, we train HUM for 50 epochs using the AdamW~\cite{loshchilov2017decoupled} optimizer with a batch size of 2 and a learning rate selected from \{1e-5, 2e-5, 5e-5\}. During training, we follow~\cite{tang2023one} to assign each sequence with 10 negative item samples. To prevent overfitting, we employ an early stopping strategy with the patience of 5 validation iterations. We set the timestamp $t$ of domain importance calculation as 50 steps for six-domain dataset. Negative samples are selected from items in the target domain within the training set, in accordance with our data partitioning principles.



\subsection{Overall Performance(RQ1)}

The results of the baselines and HUM on dataset are presented in Table~\ref{tab:performance_comparison}, from which we have the following observations:

\begin{itemize}[leftmargin=*]
    \item Among all heterogeneous user modeling baselines, semantic-based modeling methods (Recformer and LLM-Rec) outperform ID-based modeling methods (SyNCRec and UniCDR).
    This is attributed to two reasons:
    1) The incorporation of semantics through descriptive information about users and items, which enriches their representations, particularly in scenarios with numerous long-tail items and sparse interactions, enabling more accurate recommendations.
    2) Semantic-based modeling methods typically employ larger models for user representation, thereby taking advantage of the enhanced generalization capabilities of large-scale LLMs~\cite{lin2024bridging}, particularly in handling diverse data distributions.
    \item In semantic-based modeling methods, an interesting phenomenon is that the generative paradigm (\ie IDGenRec) performs worse than the discriminative paradigm (\ie Recformer). This may be due to:
    1) The multi-domain environment making item identifier representations more heterogeneous and complex, increasing the difficulty for the model to extract user preferences from sequences and generate the target item.
    2) The generative approach relying heavily on decoding to produce item identifiers, which amplifies exposure bias and error accumulation as interaction data becomes more heterogeneous. In such cases, small prediction errors in early decoding steps can propagate and compound, leading to a cascading effect that significantly degrades recommendation accuracy compared to discriminative methods.
    \item HUM exhibits improvements for all baselines across six domains in most cases, validating the effectiveness of our method. 
    The superior performance is attributed to: 
    1) The strong heterogeneous compression enhancement of LLMs, which prompts them to understand and compress heterogeneous user interactions into the user token, triggers more effective knowledge extraction from multiple domains via the masking mechanism, and encourages noise-resilient learning from heterogeneous interactions to generate unbiased user representations (refer to Section~\ref{sec:noise} for empirical evidence).
    2) The improved robustness of domain performance, which is achieved by regularizing the optimization process of each domain, thereby alleviating domain seesaw phenomenon caused by imbalanced data distribution (refer to Section~\ref{sec:robustness} for detailed analysis).
\end{itemize}

\subsection{In-depth Analysis}

\subsubsection{\textbf{Ablation Study (RQ2)}} 
\label{ablation_sec}

\begin{table*}[]
\setlength{\abovecaptionskip}{0cm}
\setlength{\belowcaptionskip}{0cm}
\caption{Ablation study of HUM. The highest score is in bold, while the second-highest is underlined.}
\label{tab:ablation_study}
\setlength{\tabcolsep}{1.9mm}{
\resizebox{0.98\textwidth}{!}{
\begin{tabular}{l|rr|rr|rr|rr|rr|rr}
\toprule
\multicolumn{1}{c|}{\textbf{}}             & \multicolumn{2}{c|}{\textbf{Scientific}}                               & \multicolumn{2}{c|}{\textbf{Office}}                                   & \multicolumn{2}{c|}{\textbf{Books}}                                    & \multicolumn{2}{c|}{\textbf{CDs}}                                      & \multicolumn{2}{c|}{\textbf{Auto}}                                     & \multicolumn{2}{c}{\textbf{Tools}}                                    \\
\multicolumn{1}{c|}{\textbf{Variants}}     & \multicolumn{1}{c}{\textbf{R@10}} & \multicolumn{1}{c|}{\textbf{N@10}} & \multicolumn{1}{c}{\textbf{R@10}} & \multicolumn{1}{c|}{\textbf{N@10}} & \multicolumn{1}{c}{\textbf{R@10}} & \multicolumn{1}{c|}{\textbf{N@10}} & \multicolumn{1}{c}{\textbf{R@10}} & \multicolumn{1}{c|}{\textbf{N@10}} & \multicolumn{1}{c}{\textbf{R@10}} & \multicolumn{1}{c|}{\textbf{N@10}} & \multicolumn{1}{c}{\textbf{R@10}} & \multicolumn{1}{c}{\textbf{N@10}} \\ \midrule\midrule
\textbf{(0): HUM}                          & { \textbf{0.0246}}                      & {\textbf{ 0.0119}}                       & \textbf{0.0155}                   & \textbf{0.0074}                         & \textbf{0.0667}                   & \textbf{0.0322}                    & \textbf{0.0161}                   & {\underline { 0.0069}}                       & {\underline { 0.0153}}                            & {\underline { 0.0067}}                             & 0.0098                            & 0.0044                            \\
\textbf{(1): HUM w/o prompt}               & 0.0223                            & 0.0102                             & 0.0126                            & 0.0058                             & 0.0595                            & 0.0297                             & 0.0105                            & 0.0045                             & 0.0129                            & 0.0059                             & 0.0097                            & 0.0048                            \\
\textbf{(2): HUM w/o user token}           & 0.0236                            & {\underline { 0.0116}}                             & 0.0143                            & {\underline { 0.0068}}                              & 0.0539                            & 0.0265                             & 0.0149                            & 0.0068                            &  \textbf{0.0157}                   & \textbf{0.0073}                       & {\underline { 0.0104}}                            & {\underline { 0.0051}}                            \\
\textbf{(3): HUM w/o user token \& prompt} & 0.0217                            & 0.0106                             & 0.0129                            & 0.0063                             & 0.0563                            & 0.0259                             & 0.0136                            & 0.0067                             & 0.0146                            & {\underline { 0.0067}}                             & 0.0091                            & 0.0042                            \\
\textbf{(4): HUM w/o mask}                 & {\underline {0.0239}}                            & 0.0113                             & {\underline { 0.0148}}                      & {\underline {0.0068}}                       & \textbf{0.0667}                   & {\underline { 0.0309}}                       & {\underline { 0.0155}}                            & \textbf{0.0072}                    & 0.0141                            & 0.0063                             &  \textbf{0.0113}                     & \textbf{0.0054}                      \\
\bottomrule
\end{tabular}
}}

\end{table*}



To thoroughly investigate the contribution of each component in HUM, we conduct an ablation study by removing each component individually (note that the ablation and analysis of the Domain Importance score are provided in Section~\ref{sec:robustness} for better illustration) and evaluating the following variants:
(0) \underline{``HUM''}, our proposed method, which enhances heterogeneous user modeling with compression enhancer and robustness enhancer.
(1) \underline{``HUM w/o prompt''}, which removes the compression prompt from HUM.  
(2) \underline{``HUM w/o user token''}, which removes the user token from HUM (\ie use \texttt{[EOS]} for representation).
(3) \underline{``HUM w/o prompt \& user token''}, which removes both the compression prompt and the tailored user token for compression.
(4) \underline{``HUM w/o mask''}, which removes the masking mechanism from HUM.
Results on six-domain dataset are depicted in Table~\ref{tab:ablation_study}. From that table, we have several key observations:
1) The inclusion of the compression prompt consistently boosts performance. This highlights the effectiveness of prompts in explicitly guiding LLMs to compress heterogeneous user behaviors, helping the model focus on useful information.
2) The user token improves model performance in most cases, mainly because it serves as a unique compression anchor for user representation, which remains unaffected by noisy information. 
3) The masking mechanism enhances model performance, demonstrating its ability to identify useful information from noisy interactions, thereby capturing signals that are more relevant to the target recommendation task.

\subsubsection{\textbf{Compression under heterogeneity (RQ2)}} 

To validate the effectiveness of compression enhancer in modeling heterogeneous user sequences, we design three experiments for analysis.

\begin{figure}[]
\setlength{\abovecaptionskip}{0.1cm}
\setlength{\belowcaptionskip}{-0.2cm}
\centering
\includegraphics[scale=0.37]{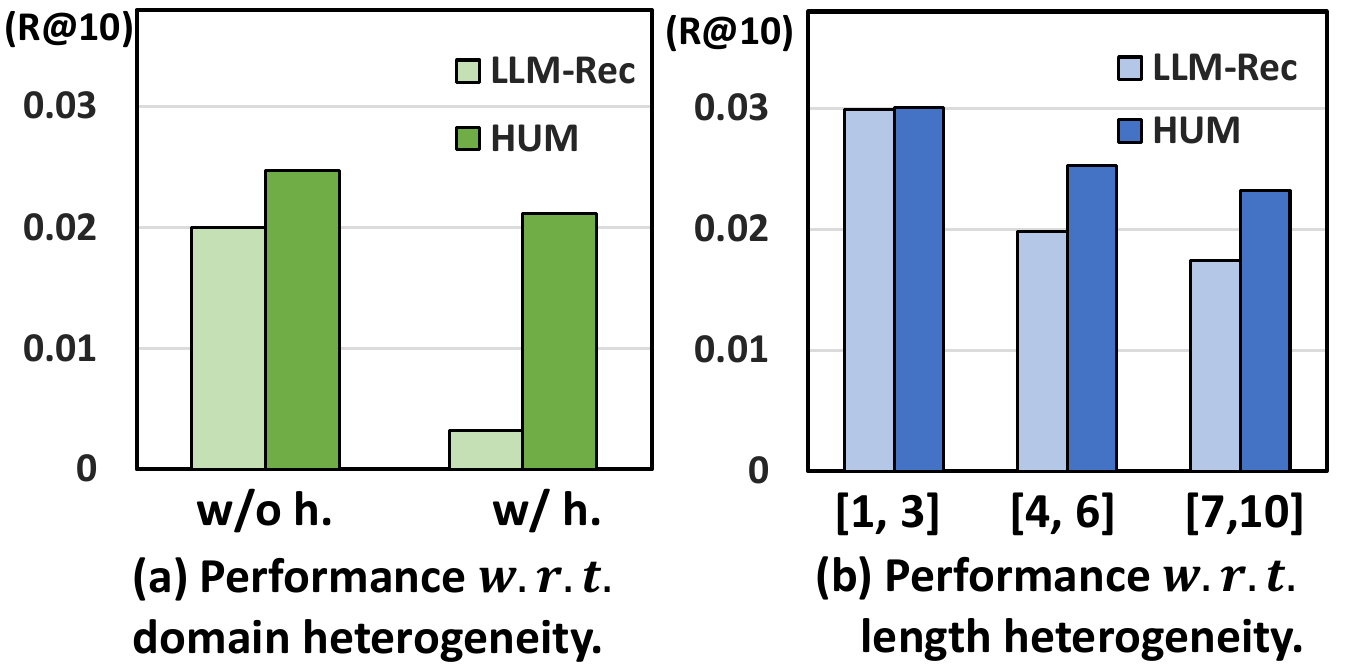}
\caption{Compression effectiveness of HUM under domain and length heterogeneity (where ``w/o h.'' denotes user sequences without multi-domain interactions).} 
\label{fig:aux-exp-compression}
\vspace{-0.0cm}
\end{figure}

\noindent$\bullet\quad$\textbf{Domain heterogeneity experiment}. 
We aim to evaluate the performance of HUM under domain-level heterogeneity, which refers to whether the items within a sequence originate from a single domain or multiple domains. 
Specifically, we divide the test data into two groups:
(0) \underline{``w/o h.''}, indicates that the user sequences do not contain interactions from other domains.
(1) \underline{``w/ h.''}, indicates that the user sequences contain interactions from other domains.
We evaluate both LLM-Rec and HUM on these two groups separately. The average results across the six domains are shown in Figure~\ref{fig:aux-exp-compression}(a). From the results, we observe the following:
1) Regardless of whether the sequences contain multi-domain interactions, HUM consistently outperforms LLM-Rec.
2) On sequences with multi-domain interactions, HUM shows a substantial improvement over LLM-Rec, which further validates HUM’s strong performance under domain-level heterogeneity and further highlight its advantage in compressing heterogeneous user sequences and capturing users’ diverse preferences.

\noindent$\bullet\quad$\textbf{Length heterogeneity experiment}. 
We further evaluate the performance of HUM under sequential heterogeneity, which refers to differences in sequence length, with longer sequences generally implying greater diversity and complexity.
Specifically, we divide the test data into three groups based on the number of items in each sequence:
(0) \underline{``[1,3]''}, indicates sequences with a length between 1 and 3.
(1) \underline{``[4,6]''}, indicates sequences with a length between 4 and 6;
(2) \underline{``[7,10]''}, indicates sequences with a length between 7 and 10.
Similarly, we evaluate both LLM-Rec and HUM on these three groups separately. The average results across the six domains are shown in Figure~\ref{fig:aux-exp-compression}(b). From the results, we observe the following:
1) HUM consistently achieves higher performance than LLM-Rec across all length groups.
2) Compared to LLM-Rec, HUM delivers more stable results in groups with longer sequences, demonstrating HUM’s effectiveness in handling length-based heterogeneity.

\noindent$\bullet\quad$\textbf{Compression attention experiment}. 
We aim to investigate the impact of attention design in the compression module, which is studied by comparing unidirectional attention with bidirectional attention.
Specifically, we consider the following two variants for comparison:
(0) \underline{``HUM w/ bidirectional atten.''}, which replaces the unidirectional attention in HUM with bidirectional attention;
(1) \underline{``HUM''}, our proposed method.
From Table~\ref{tab:aux-exp-atten}, we can observe that unidirectional attention achieves better performance than bidirectional one, demonstrating the advantage of our design in effectively condensing user interactions and capturing diverse preferences.

\begin{table}[t]
\small
\caption{Performance comparison between HUM w/ bidirectional attention and HUM. The highest score is in bold.}
\label{tab:aux-exp-atten}
\setlength{\tabcolsep}{1.5mm}
\begin{tabular}{ccccccccc}
\toprule
\textbf{Method} & \textbf{R@5} & \textbf{R@10} & \textbf{N@5} & \textbf{N@10} \\
\midrule
HUM w/ bidirectional atten. & 0.0119 & 0.0211 & 0.0067 & 0.0097 \\
\textbf{HUM} & \textbf{0.0145} & \textbf{0.0247} & \textbf{0.0085} & \textbf{0.0116} \\
\bottomrule
\end{tabular}
\vspace{-0.3cm}
\end{table}

\subsubsection{\textbf{Robustness in seen domains (RQ2)}} 

\label{sec:robustness}

\begin{figure}[]
\setlength{\abovecaptionskip}{0.1cm}
\setlength{\belowcaptionskip}{-0.2cm}
\centering
\includegraphics[scale=0.47]{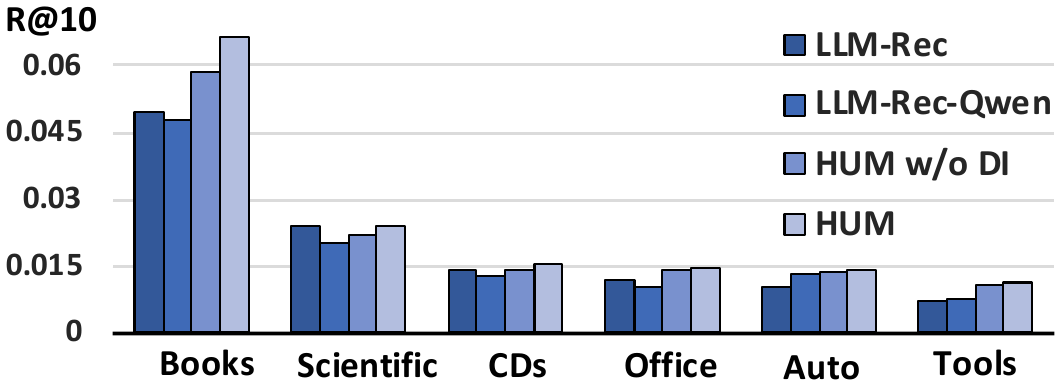}
\caption{Domain robustness performance on a six-domain dataset using four variants: 0) \underline{``LLM-Rec''}: a semantic-based modeling method. 1) \underline{``LLM-Rec-Qwen''}: using LLMs as the backbone from 0). 2) \underline{``HUM w/o DI''}: HUM without domain importance score. 3) \underline{``HUM''}: our proposed method. }
\label{fig:exp-robustness}
\vspace{-0.3cm}
\end{figure}


To verify whether HUM mitigate the domain seesaw phenomenon (refer to Figure~\ref{fig:domain_seesaw}) and achieve robust performance across domains, we compare the following methods: 
(0) \underline{``LLM-Rec''}, one of the strongest semantic-based modeling methods.
(1) \underline{``LLM-Rec-Qwen''}, a variant of LLM-Rec using a decoder-only LLM, \ie Qwen2.5-1.5b as backbone.
(2) \underline{``HUM w/o DI''}, HUM without domain importance score.
(3) \underline{``HUM''}, our proposed method.
The results are presented in Figure~\ref{fig:exp-robustness}. We can find that:
1) On the six-domain dataset, directly applying a strong LLM (\ie LLM-Rec-Qwen) results in domain seesaw phenomenon, where gains in some domains (\eg Books) come at the cost of other (\eg Scientific), highlighting the need for optimization constraints to ensure robustness.
2) The domain importance score improves performance across domains, confirming its role as an effective regularizer that promotes balanced optimization.
\subsubsection{\textbf{Noise resistance(RQ2)}} 
\label{sec:noise}

\begin{figure}[]
\setlength{\abovecaptionskip}{0.1cm}
\setlength{\belowcaptionskip}{-0.1cm}
\centering
\includegraphics[scale=0.35]{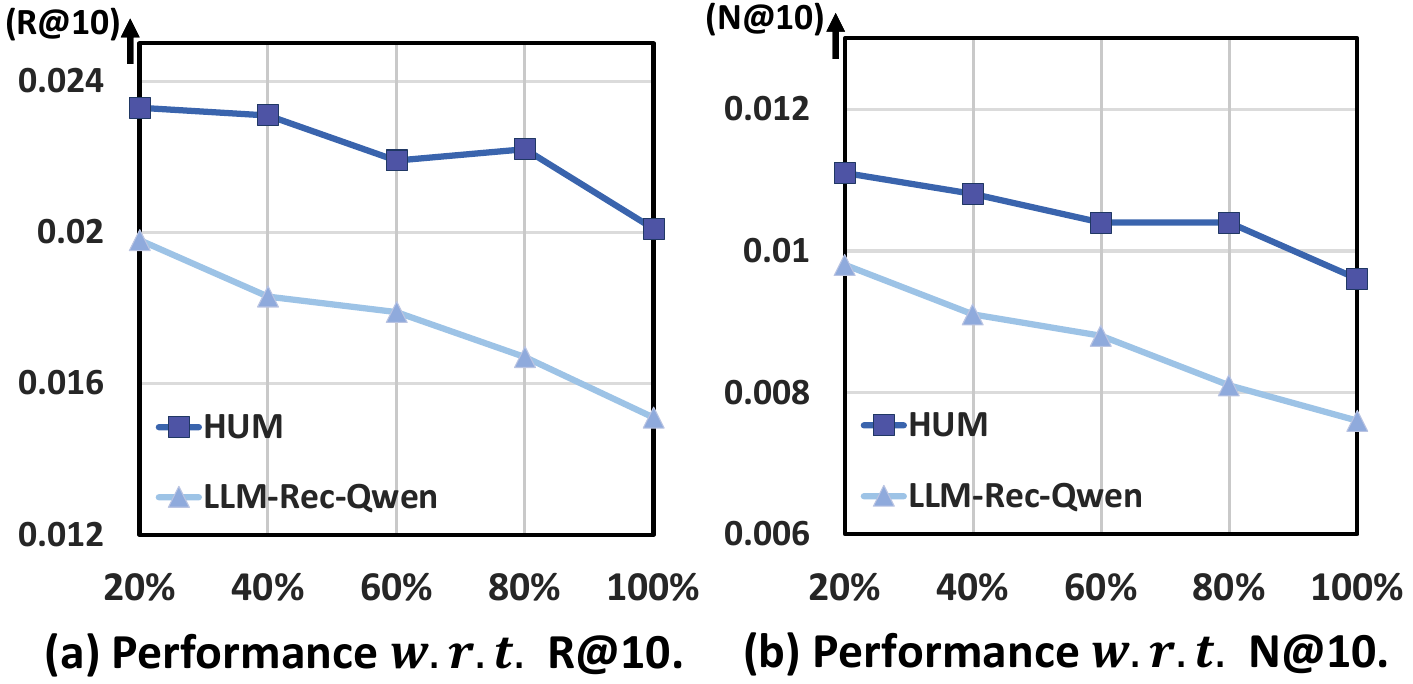}
\caption{The illustration of strong noise resistance of HUM.}
\label{fig:exp-noisy}
\vspace{-0.3cm}
\end{figure}

In multi-domain recommendation, users' heterogeneous interactions often contain noise, which hinders accurate preference modeling. To evaluate whether HUM can infer true user representations from noisy data, we compare it with the pure LLM-based method, \ie LLM-Rec-Qwen. Specifically, we randomly select a proportion of users, ranging from 20\% to 100\%, and replace a subset of their interacted items with non-interacted items as random noise. Figure~\ref{fig:exp-noisy} shows the results on the Scientific domain, with similar trends observed in other domains.
From the results, we observe the following:
1) As the proportion of users with noisy interactions increases from 20\% to 100\%, the performance of both HUM and LLM-Rec-Qwen gradually declines. This is expected, as capturing user preferences becomes more challenging with increased noise.
2) HUM consistently outperforms LLM-Rec-Qwen even under a high level of noise, demonstrating its strong resistance against random noise. This is attributed to HUM’s ability to effectively integrate and compress heterogeneous interactions, mitigating the impact of noise.
In contrast, 3) LLM-Rec-Qwen is more vulnerable to noisy interactions, as it tends to amplify the negative effects of noise by merely interpreting the semantics of noisy tokens, ultimately leading to poor performance.

\subsubsection{\textbf{Generalization to unseen domains (RQ3)}} \label{sec: generalization}

\begin{table}[]
\setlength{\abovecaptionskip}{0.3cm}
\setlength{\belowcaptionskip}{0.2cm}
\caption{Comparison of generalization performance on four cold domains using four variants: 0) \underline{``LLM-Rec''}: a semantic-based modeling method. 1) \underline{``LLM-Rec-Qwen''}: using LLMs as the backbone from 0). 2) \underline{``HUM''}: our proposed method. 3) \underline{``HUM w/o u.t.''}: HUM without user token.}
\label{tab:generalization}
\setlength{\tabcolsep}{1.2mm}{
\resizebox{0.44\textwidth}{!}{
\begin{tabular}{c|ccccc}
\toprule
\textbf{Dataset}                      & \multicolumn{1}{c}{\textbf{Metric}} & \textbf{LLM-Rec} & \textbf{\begin{tabular}[c]{@{}c@{}}LLM-Rec\\ -Qwen\end{tabular}} & \textbf{HUM}    & \textbf{\begin{tabular}[c]{@{}c@{}}HUM \\ w/o u.t.\end{tabular}} \\ \midrule
\multirow{4}{*}{\textbf{Instruments}} & \textbf{R@5}                         & 0.0052           & 0.0052                                                           & \textbf{0.0080} & {\underline { 0.0066}}                                                     \\
                                      & \textbf{N@5}                         & 0.0096           & 0.0094                                                           & \textbf{0.0142} & {\underline { 0.0120}}                                                     \\
                                      & \textbf{R@10}                        & 0.0026           & 0.0030                                                           & \textbf{0.0050} & {\underline { 0.0036}}                                                     \\
                                      & \textbf{N@10}                        & 0.0041           & 0.0043                                                           & \textbf{0.0070} & {\underline { 0.0054}}                                                     \\ \midrule
\multirow{4}{*}{\textbf{Games}}       & \textbf{R@5}                         & 0.0093           & 0.0126                                                           & \textbf{0.0161} & {\underline { 0.0128}}                                                     \\
                                      & \textbf{N@5}                         & 0.0178           & {\underline { 0.0218}}                                                     & \textbf{0.0237} & 0.0207                                                           \\
                                      & \textbf{R@10}                        & 0.0053           & 0.0074                                                           & \textbf{0.0094} & {\underline { 0.0078}}                                                     \\
                                      & \textbf{N@10}                        & 0.0080           & {\underline { 0.0105}}                                                     & \textbf{0.0118} & 0.0103                                                           \\ \midrule
\multirow{4}{*}{\textbf{Arts}}        & \textbf{R@5}                         & 0.0051           & 0.0053                                                           & {\underline { 0.0060}}     & \textbf{0.0066}                                                  \\
                                      & \textbf{N@5}                         & 0.0102           & \textbf{0.0116}                                                  & 0.0110           & {\underline { 0.0112}}                                                     \\
                                      & \textbf{R@10}                        & 0.0028           & 0.0031                                                           & {\underline { 0.0035}}    & \textbf{0.0036}                                                  \\
                                      & \textbf{N@10}                        & 0.0044           & \textbf{0.0051}                                                  & \textbf{0.0051} & \textbf{0.0051}                                                  \\ \midrule
\multirow{4}{*}{\textbf{Sports}}      & \textbf{R@5}                         & 0.0066           & 0.0076                                                           & \textbf{0.0100} & {\underline { 0.0079}}                                                     \\
                                      & \textbf{N@5}                         & 0.0110            & 0.0117                                                           & \textbf{0.0140}  & \textbf{0.0140}                                                   \\
                                      & \textbf{R@10}                        & 0.0039           & 0.0042                                                           & \textbf{0.0061} & {\underline { 0.0043}}                                                     \\
                                      & \textbf{N@10}                        & 0.0053           & 0.0055                                                           & \textbf{0.0074} & {\underline { 0.0063}}                                                     \\ \bottomrule
\end{tabular}
}}
\label{tab:generalization}
\end{table}


Open-domain recommendation~\cite{10.1007/978-3-540-88192-6_31} extends beyond multi-domain recommendation by involving more domains and greater heterogeneity in interactions, posing higher demands on generalization.
To evaluate the generalization ability of HUM, we compare the following variants:
(0) \underline{``LLM-Rec''}, one of the strongest semantic-based modeling methods.
(1) \underline{``LLM-Rec-Qwen''}, a variant of LLM-Rec using a decoder-only LLM, \ie Qwen2.5-1.5b as backbone.
(2) \underline{``HUM''}, our proposed method with compression enhancer and robustness enhancer.
(3) \underline{``HUM w/o u.t.''}, a variant of HUM without the user token.
We test these models on four cold domains: ``Instruments'', ``Games'', ``Arts'' and ``Sports'' from the latest Amazon dataset~\cite{hou2024bridging}, sampling 10,000 users per domain for evaluation. 
From the Table~\ref{tab:generalization}, we have several key observations:
1) HUM achieves the best performance across domains, demonstrating strong generalization. This shows the effectiveness of the user token as a compression carrier, which not only delivers strong results in seen domains but also impacts unseen domains.
2) LLM-Rec-Qwen slightly outperforms LLM-Rec, indicating that stronger LLMs can enhance generalization performance. Combined with HUM’s results, this highlights that effective knowledge compression, rather than model size alone, is crucial for generalization in open-domain recommendation.

\subsubsection{\textbf{Scalability under heterogeneity (RQ3)}} 

\begin{figure}[]
\setlength{\abovecaptionskip}{0.1cm}
\setlength{\belowcaptionskip}{0cm}
\centering
\includegraphics[scale=0.35]{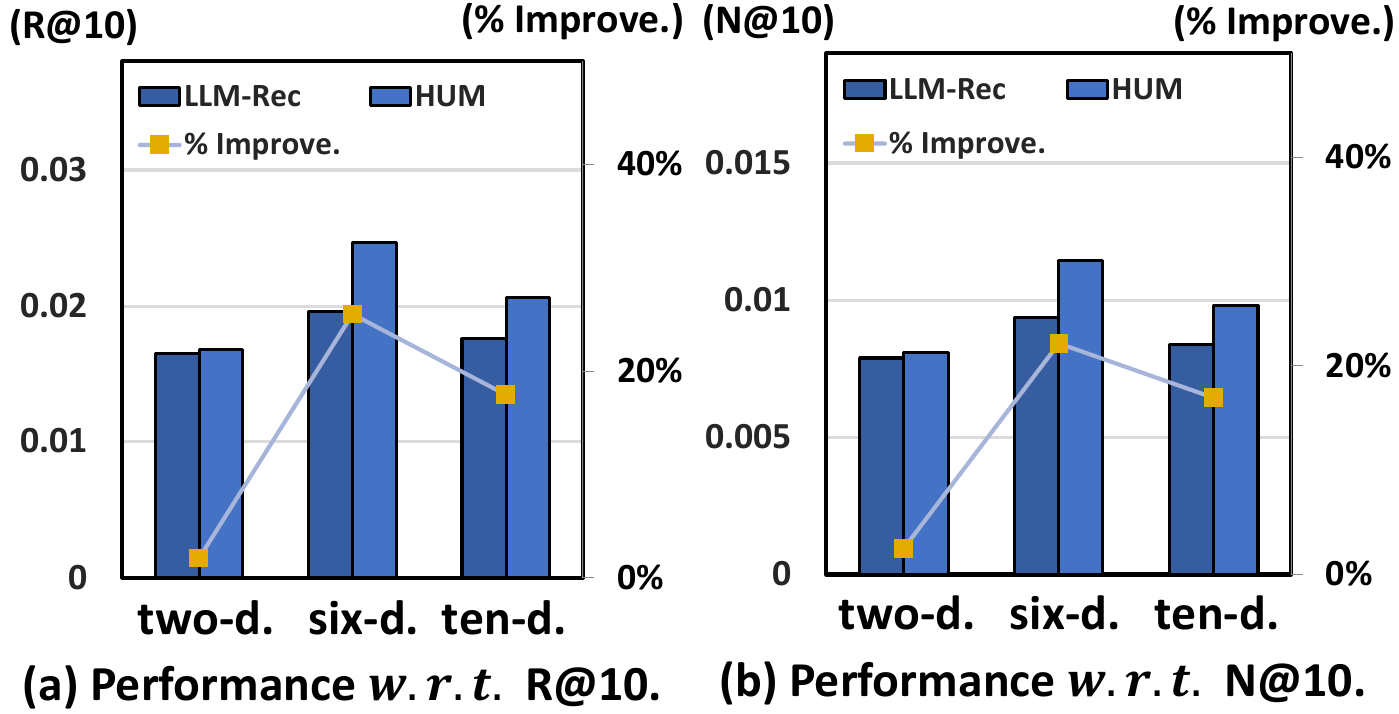}
\caption{The illustration of scalability across different levels of heterogeneity (\ie two domain, six domains, and ten domains, abbreviated as ``two-d.'', ``six-d.'', and ``ten-d.'').}
\label{fig:exp-scale}
\vspace{-0.3cm}
\end{figure}

In open-domain recommendation, increasing data scale leads to greater heterogeneity in user representations~\cite{iacob2024deptdecoupledembeddingspretraining}. To investigate whether HUM can achieve strong performance under varying degrees of heterogeneity, we evaluate both LLM-Rec and HUM on three datasets of different scales: a two-domain version, a six-domain version, and a ten-domain version. Results are shown in Figure~\ref{fig:exp-scale}, with performance averaged within each scale group. From Figure~\ref{fig:exp-scale}, we observe:
1) HUM consistently outperforms LLM-Rec across all levels of heterogeneity, demonstrating its  ability to handle diverse user interactions. This highlights the value of enhancing LLMs with compression and robustness mechanisms.
2) The performance gap between HUM and LLM-Rec widens as heterogeneity increases (\eg in six-domain and ten-domain settings), indicating HUM’s effectiveness in modeling rich, heterogeneous behaviors in open-domain environments.

\subsubsection{\textbf{Effect of mask ratio $r$}} 
\label{sec:mask_ratio}

\begin{figure}[t]
\setlength{\abovecaptionskip}{0.05cm}
\setlength{\belowcaptionskip}{-0.2cm}
\centering
\includegraphics[scale=0.26]{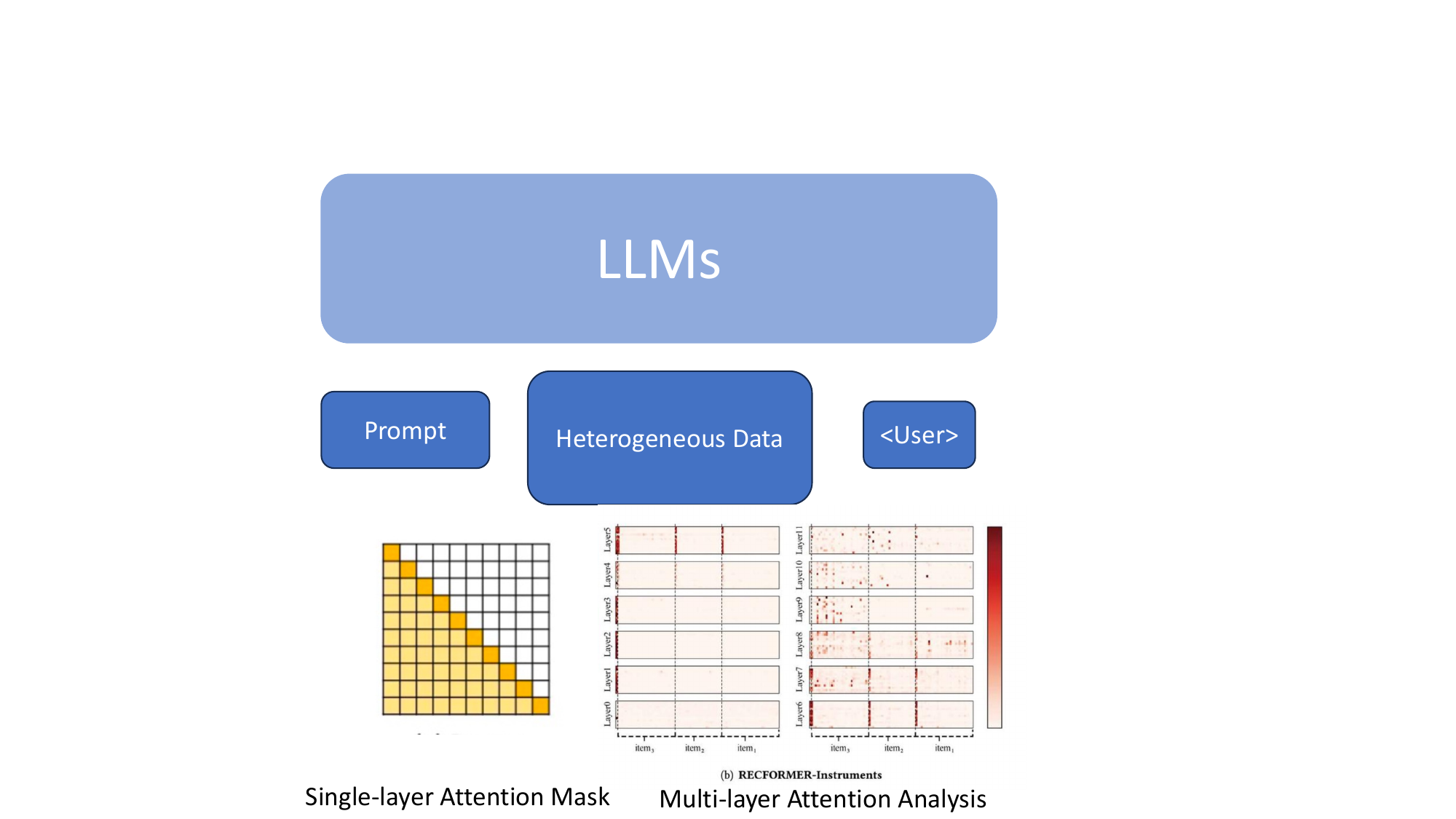}
\caption{Performance of HUM with different $r$.}
\label{fig:exp-mask}
\vspace{-0.3cm}
\end{figure}

To investigate the impact of the mask ratio $r$ on the accuracy of HUM,  we vary the ratio ${r}$ from $10\%$ to $40\%$ and present the results in Figure~\ref{fig:exp-mask}. 
It is observed that
1) Moderately increasing $r$ can improve performance, as the masking mechanism helps the model leverage useful heterogeneous information from other domains while filtering noisy interactions, enhancing recommendation in the target domain.
2) However, excessively increasing $r$ may inversely hurt performance. This could be because an overly high mask ratio leads to information loss, resulting in suboptimal learning of target domain information.
3) These observations suggest that there is a sweet spot for $r$, where the benefits of cross-domain information outweigh the costs of missing target domain details.

%% file: 4_related_work.tex
\section{Related Work}
\label{sec:related_work}



\vspace{2pt} \noindent$\bullet\quad$\textbf{LLM-based Recommendation.}
Large Language Models (LLMs) have shown exceptional success in various recommendation tasks~\cite{wang2024learnableitemtokenizationgenerative, fu2023unified, ma2024xrec, bao2023tallrec, deng2025cram, zhao2025federated, xu2025personalizedgenerationlargemodel}.
There are two primary methods for using LLMs in recommendation:
1) LLM-based recommender, that directly employs LLMs as the recommendation models~\cite{li2023e4srec, zhang2023collm,lin2023clickprompt, bao2023tallrec,lin2023rella,prakash2023cr, zheng2023adapting, lin2025order, Li2024ASO}. 
2) LLM-enhanced recommender, while utilize LLMs for data augmentation~\cite{wei2024llmrec, xi2023towards,liu2024once} and representation~\cite{qiu2021u,ren2023representation,Li2023TextIA, tang2023one}. 
Leveraging LLMs for representation refers to generating embedding for entities(\eg users or items) through LLMs. 
For instance, LLM-Rec~\cite{tang2023one} uses BERT~\cite{devlin2018bert}, OPT~\cite{zhang2022opt} and Flan-T5~\cite{chung2024scaling} to create user and item embedding based on their semantic information(\ie plain text).
In this work, we focus on leveraging LLMs for representing heterogeneous user interactions, thus we do not dive deeper into the discussion of representation methods in other fields of recommendation.

\vspace{2pt} \noindent$\bullet\quad$\textbf{Multi-domain Recommendation.}
Multi-domain recommendation~\cite{10.1145/3459637.3481941, tang2023one, fu2023unified, park2024pacer} models user preferences based on their heterogeneous interactions across multiple domains, encompassing cross-domain recommendation~\cite{ijcai2017p343, cao2022c2dsr, cao2023unicdr, 10.1145/3589334.3645351} and transferable recommendation~\cite{hou2022unisrec, Li2023TextIA, tan2024towards}.
Previous heterogeneous user modeling methods in multi-domain recommendation can be broadly categorized into two approaches:  
1) ID-based modeling~\cite{cao2022c2dsr, cao2023unicdr, 10.1145/3589334.3645351, park2024pacer}, which utilizes unique IDs to index embeddings for user representations and employs specific mechanisms to optimize them.  
For instance, UniCDR~\cite{cao2023unicdr} applies specialized aggregators (\eg user-attention and item-similarity) to capture both domain-shared and domain-specific representations. 
However, this approach heavily depends on sufficient interaction data, making it difficult to generalize to new scenarios and leading to suboptimal performance in long-tail domains. 2) Semantic-based modeling~\cite{hou2022unisrec, Li2023TextIA, tang2023one}, which leverages the semantic information of user-interacted items to generate user representations.  
Recformer~\cite{Li2023TextIA} and LLM-Rec~\cite{tang2023one} both use textual descriptions of users' interacted items to construct user representations and employ token-level user attention to model preferences.  
However, these methods focus on integrating the semantics of heterogeneous tokens while neglecting unbiased compression of noisy interactions in user behaviors, which weakens their resistance to noise and hinders accurate preference modeling, contributing to the domain seesaw phenomenon.

%% file: 5_conclusion.tex
\section{Conclusion and Future Work}
\label{sec:conclusion}

In this study, we conducted a comprehensive analysis of the optimal features for heterogeneous user modeling in LLM-based open-domain recommendation. We then introduce HUM, which incorporates a compression enhancer and a robustness enhancer to achieve strong compression capabilities for LLMs and enhanced robustness across different domains. Extensive experiments on heterogeneous datasets demonstrate the superiority of HUM, achieving excellent robustness on seen domains, strong noise resistance, high generalization on unseen domains, and scalability in heterogeneous user modeling for LLM-based open-domain recommendation.

This work emphasizes the importance of heterogeneous user modeling in LLM-based open-domain recommendation and suggests several promising directions for future research:
1) Future research may investigate theoretical foundations for compression mechanisms in heterogeneous user modeling, to better understand and justify their effectiveness from an analytical perspective.
2) Designing compression-aware architectures for large language models presents an important direction, aiming to reduce inference overhead while preserving the ability to model heterogeneous user information effectively.
3) Another direction is to explore how to combine heterogeneous user modeling with generative recommendation, enabling the model to understand heterogeneous interactions and generate target items across multiple domains.